\documentclass[twocolumn, 9pt]{extarticle}
\usepackage[a4paper, margin=1.8cm]{geometry}

\pdfoutput=1

\usepackage[english]{babel}
\usepackage{subfig}
\usepackage{lipsum}
\usepackage{xcolor} 
\usepackage{amsmath,amsfonts,amssymb,amsthm} 
\usepackage{calc} 
\usepackage[utf8]{inputenc} 
\usepackage{helvet} 
\usepackage{textcomp}
\usepackage{fancyhdr} 
\usepackage{nicefrac}
\usepackage{multicol}
\usepackage{titlesec}
\usepackage{hyperref}
\usepackage{needspace}
\usepackage{tikz}
\usepackage{pgfplots}
\usepackage{fontawesome}
\usepackage{quoting}
\usepackage{csquotes}
\usepackage{multirow}
\usepackage{mathtools}
\usepackage{float}
\usepackage{xfrac}

\include{commands}

\definecolor{fogliadite}{RGB}{0,128,128} 
\definecolor{fogliaditeScuro}{RGB}{0,80,80} 
\definecolor{lightblue}{RGB}{00, 123, 255}
\definecolor{warning}{RGB}{229, 162, 41}
\definecolor{darkgray}{RGB}{80, 80, 80}

\SetBlockThreshold{2}
\newcommand\myblockquote[1]{\vspace*{-10pt}
  \blockquote{\hspace*{1em}\emph{``#1''}}\par}





\hypersetup{
     colorlinks   = true,
     citecolor    = gray,
     linkcolor=black
}

\begin{document}
\title{\Huge{Mitigating TLS compromise with ECDHE and SRP}}

\author{Aron Wussler \\Proton Technologies AG \\ \texttt{aron@wussler.it}}
\maketitle

\begin{abstract}
  The paper reviews an implementation of an additional encrypted tunnel within TLS to further secure and authenticate the traffic of personal information between ProtonMail's frontends and the backend, implementing its key exchange, symmetric packet encryption, and validation.
  Technologies such as \emph{Secure Remote Password} (SRP) and the \emph{Elliptic Curves Diffie Hellman Ephemeral} (ECDHE) exchange are used for the key exchange, verifying the public parameters through PGP signatures.
  The data is then transferred encrypted with AES-128-GCM.

  This project is meant to integrate TLS security for high security data transfer, offering a flexible model that is easy to implement in the frontends by reusing part of the standard already existing in the PGP libraries.
\end{abstract}

\section{Introduction}
\textit{Transport Layer Security} (TLS) is the protocol that secures most of our web traffic every day, while connecting to most of the websites we visit.
It is often represented as a green lock next to the address of the webpage, this means the connection is encrypted and the content is loaded securely.
The security of the connection is provided by a cipher suite, made up of an asymmetric encryption algorithm to exchange a key, and a symmetric one to transfer encrypted packets with the mediated key.
The authenticity of the connection is guaranteed by a certificate, released by an authority that certifies the ownership of the domain.

This does not fully protect the user from an attack, several cases of TLS compromise are known \cite{KCI-attack}, both malicious \cite{ca-abuse, MD5-collision, weak-cipher, BREACH} and consensual \cite{tls-proxy}, predating on the naivety of a user or in corporate-issued devices.
The most common situation is the inclusion of an additional (malicious) authority that is now allowed to issue rogue certificates and mark every connection as protected, defeating the purpose of TLS.

A high-security service, such as ProtonMail, is interested in adding a further layer of protection to the connection negotiating an encrypted tunnel within TLS, with its own certification system, key exchange, and symmetric encryption.
The implementation of this tunnel is fully in the application layer, thus not dependent on the browser or the underlying operating system.

All the frontends connecting to ProtonMail's backend API will have to negotiate a session key as the first interaction, to be then used to encrypt every packet.
While generally using tunnels within tunnels is frowned upon, because of the performance drawbacks compared to the slight security increase, this might be a useful niche application: the larger issue here is not the encryption scheme, but rather the certification system.

The key exchange leverages elliptic curves, further secured with a byproduct of the password exchange when logging in.
To ensure the authenticity of this procedure, the public parameters of the exchange are signed with a PGP-key stored in ProtonMail's key verification system, based on the Merkle tree technology.
This project, matched with proper source code transparency and verification, can easily prove the authenticity of the connection without compromises or intermediaries.

\section{Design}
\subsection{Threat Model}
TLS compromise is the main threat we are going to address with this countermeasure, it might seem an unlikely scenario but as we are going to se in the next paragraphs there is plenty of real life cases where this happened.
This is mostly done on puropse, by large and medium corporate environments to surveil their users' traffic, with little care about certificate validation and best security practices, like not using always the strongest ciphers available.

In other, worse cases, it is compromised maliciously, by attackers and goverment entities with various techniques, most commonly by abusing the trust chain or exploiting implementation bugs, rather than cracking cryptography.

In this section these compromises are going to be analyzed, to examine security of TLS implementation until the latest 1.3 version.

\subsubsection{TLS proxy}
TLS proxies are used in many business to spy on the user's internet traffic to avoid data leak or to filter content, or in malicious settings they are also known as KCI (\textit{Key Compromise Impersonation}) \cite{KCI-attack}.
This malicious attack is mostly performed through social engineering and convincing an unknowledgeable user to install certificates.

The compromise consists in a MITM (\textit{Man In The Middle}) attack: a root certificate is installed on the client, and when a connection to the server is done, a proxy, or TIA (\textit{TLS Intercept Application}), intercepts it to act as a middleman, negotiating a session key both, with the client and the server, thus being able to snoop in the communication.
The client displays no warning while this is happening because the chain of trust is guaranteed by the extra root certificate, hence not recognising the proxy's fake certificate as untrustworthy.

Up to TLS 1.2 \cite{RFC5246} the TIA behaviour is the following:
\begin{itemize}
  \item the TIA forwards the client's \texttt{ClientHello} message as is to the server and inspects the resulting \texttt{ServerHello} and X.509 certificates;
  \item based on the name in the certificate, the TIA may:
  \begin{itemize}
    \item ``drop out'' of the connection, i.e. allow the client and server to communicate directly;
    \item interpose itself, answering the client with an alternative \texttt{ServerHello} and the server with an alternative \texttt{ClientKeyExchange}, negotiating different encrypted sessions with each party and forwarding so that it can see the plaintext of the connection, as shown in figure \ref{fig-tls-tia-1.2-s}.
  \end{itemize}
\end{itemize}

\begin{figure}[ht]
  \centering
  \begin{tikzpicture}
    \node at (-3.0, 0) {\large C};
    \node at (0, 0) {\large TIA};
    \node at (3.0, 0) {\large S};

    \draw[line width=0.5mm] (3.0,-0.5) -- (3.0,-7.3);
    \draw[line width=0.5mm] (0,-0.5) -- (0,-7.3);
    \draw[line width=0.5mm] (-3.0,-0.5) -- (-3.0,-7.3);

    \draw[->, line width=0.3mm] (-2.8,-1) -- (-0.2,-1) node[midway, above] {\small CH};
    \draw[<-, line width=0.3mm] (-2.8,-1.7) -- (-0.2,-1.7) node[midway, above] {\small SH, SC, SKE, SHD};
    \draw[->, line width=0.3mm] (-2.8,-2.4) -- (-0.2,-2.4) node[midway, above] {\small CKE, CCS, [FIN]};
    \draw[<-, line width=0.3mm] (-2.8,-3.1) -- (-0.2,-3.1) node[midway, above] {\small CCS, [FIN]};
    \draw[<->, line width=0.3mm] (-2.8,-4.2) -- (-0.2,-4.2) node[midway, above] {\small [HTTP]};

    \draw[<-, line width=0.3mm] (2.8,-4.2) -- (0.2,-4.2) node[midway, above] {\small CH};
    \draw[->, line width=0.3mm] (2.8,-4.9) -- (0.2,-4.9) node[midway, above] {\small SH, SC, SKE, SHD};
    \draw[<-, line width=0.3mm] (2.8,-5.6) -- (0.2,-5.6) node[midway, above] {\small CKE, CCS, [FIN]};
    \draw[->, line width=0.3mm] (2.8,-6.3) -- (0.2,-6.3) node[midway, above] {\small CCS, [FIN]};
    \draw[<->, line width=0.3mm] (2.8,-7) -- (0.2,-7) node[midway, above] {\small [HTTP]};
  \end{tikzpicture}
  \caption{Asynchronous TLS 1.2 proxy. Square brackets indicate encrypted packets.
    {\color{darkgray} TIA: TLS Intercept Application, CH: ClientHello, SH: ServerHello, SC: Server X.509 certificate, SKE: Server Key Ephemeral, SHD: ServerHelloDone, CKE: Client Key Ephemeral, CCS: ChangeCipherSpec, FIN: only encrypted messages following.}}
  \label{fig-tls-tia-1.2-a}
\end{figure}
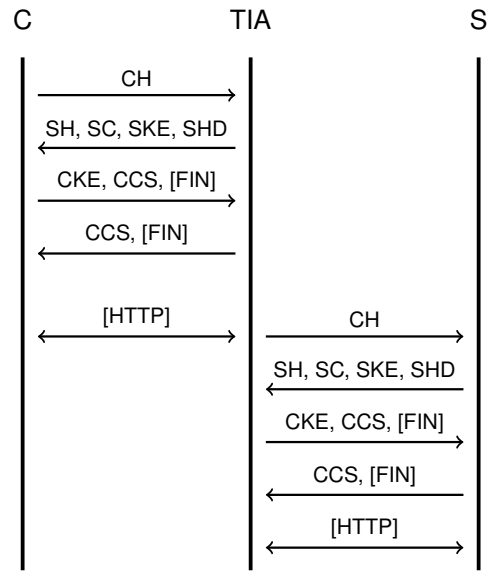

The attack can be performed also asynchronously, generating the server certificate based upon the \texttt{ClientHello}, see figure \ref{fig-tls-tia-1.2-a}.
\begin{figure}[ht]
  \centering
  \begin{tikzpicture}
    \node at (-3.0, 0) {\large C};
    \node at (0, 0) {\large TIA};
    \node at (3.0, 0) {\large S};

    \draw[line width=0.5mm] (3.0,-0.5) -- (3.0,-5.2);
    \draw[line width=0.5mm] (0,-0.5) -- (0,-5.2);
    \draw[line width=0.5mm] (-3.0,-0.5) -- (-3.0,-5.2);

    \draw[->, line width=0.3mm] (-2.8,-1) -- (-0.2,-1) node[midway, above] {\small CH};
    \draw[<-, line width=0.3mm] (-2.8,-2.1) -- (-0.2,-2.1) node[midway, above] {\small SH, SC, SKE, SHD};
    \draw[->, line width=0.3mm] (-2.8,-2.8) -- (-0.2,-2.8) node[midway, above] {\small CKE, CCS, [FIN]};
    \draw[<-, line width=0.3mm] (-2.8,-3.9) -- (-0.2,-3.9) node[midway, above] {\small CCS, [FIN]};
    \draw[<->, line width=0.3mm] (-2.8,-4.9) -- (-0.2,-4.9) node[midway, above] {\small [HTTP]};

    \draw[<-, line width=0.3mm] (2.8,-1.2) -- (0.2,-1.2) node[midway, above] {\small CH};
    \draw[->, line width=0.3mm] (2.8,-1.9) -- (0.2,-1.9) node[midway, above] {\small SH, SC, SKE, SHD};
    \draw[<-, line width=0.3mm] (2.8,-3) -- (0.2,-3) node[midway, above] {\small CKE, CCS, [FIN]};
    \draw[->, line width=0.3mm] (2.8,-3.7) -- (0.2,-3.7) node[midway, above] {\small CCS, [FIN]};
    \draw[<->, line width=0.3mm] (2.8,-4.9) -- (0.2,-4.9) node[midway, above] {\small [HTTP]};

  \end{tikzpicture}
  \caption{Inline TLS 1.2 proxy. Square brackets indicate encrypted packets.
    {\color{darkgray} TIA: TLS Intercept Application, CH: ClientHello, SH: ServerHello, SC: Server X.509 certificate, SKE: Server Key Ephemeral, SHD: ServerHelloDone, CKE: Client Key Ephemeral, CCS: ChangeCipherSpec, FIN: only encrypted messages following.}}
  \label{fig-tls-tia-1.2-s}
\end{figure}
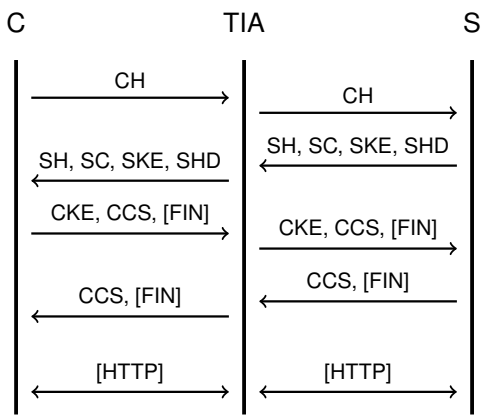

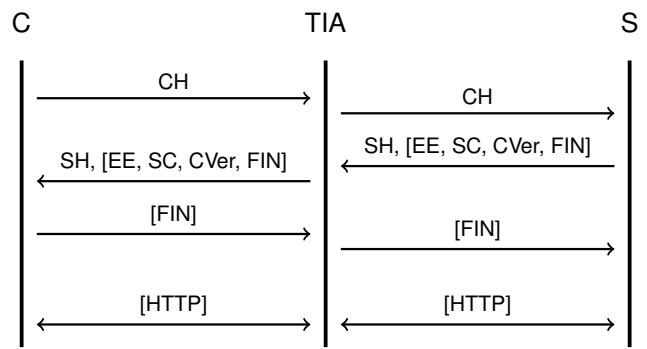
\begin{figure}[ht]
  \centering
  \begin{tikzpicture}
    \node at (-4, 0) {\large C};
    \node at (0, 0) {\large TIA};
    \node at (4, 0) {\large S};

    \draw[line width=0.5mm] (4,-0.5) -- (4,-4.3);
    \draw[line width=0.5mm] (0,-0.5) -- (0,-4.3);
    \draw[line width=0.5mm] (-4,-0.5) -- (-4,-4.3);

    \draw[->, line width=0.3mm] (-3.8,-1) -- (-0.2,-1) node[midway, above] {\small CH};
    \draw[<-, line width=0.3mm] (-3.8,-2.1) -- (-0.2,-2.1) node[midway, above] {\small SH, [EE, SC, CVer, FIN]};
    \draw[->, line width=0.3mm] (-3.8,-2.8) -- (-0.2,-2.8) node[midway, above] {\small [FIN]};
    \draw[<->, line width=0.3mm] (-3.8,-4) -- (-0.2,-4) node[midway, above] {\small [HTTP]};

    \draw[<-, line width=0.3mm] (3.8,-1.2) -- (0.2,-1.2) node[midway, above] {\small CH};
    \draw[->, line width=0.3mm] (3.8,-1.9) -- (0.2,-1.9) node[midway, above] {\small SH, [EE, SC, CVer, FIN]};
    \draw[<-, line width=0.3mm] (3.8,-3) -- (0.2,-3) node[midway, above] {\small [FIN]};
    \draw[<->, line width=0.3mm] (3.8,-4) -- (0.2,-4) node[midway, above] {\small [HTTP]};
  \end{tikzpicture}
  \caption{Inline TLS 1.3 proxy. Square brackets indicate encrypted packets.
    {\color{darkgray} TIA: TLS Intercept Application, CH: ClientHello, SH: ServerHello, EE: Encrypted Extension, SC: Server X.509 certificate, CVer: Certificate Verify, FIN: only encrypted messages following.}}
  \label{fig-tls-tia-1.3}
\end{figure}

It is a common misconception that TLS 1.3 \cite{RFC8446} is immune to this attack, but in reality it will make it just harder and more computationally expensive.
The main difference between 1.2 and 1.3 is the deprecation of the use of RSA key exchange in favor of (EC)DHE, which implies that, for practical purposes, a TIA must be inline to participate in the TLS handshake, see figure \ref{fig-tls-tia-1.3}.

This makes the ``drop out'' of the encrypted session impossible, i.e. a TIA is unable to selectively choose which connections to monitor: once you proxy a connection, you have to proxy it until it’s done.
Furthermore, the version 1.3 prevents downgrade attacks, disallowing a TIA that only supports TLS 1.2 to negotiate a TLS 1.3 connection, without triggering an \texttt{illegal\_parameter} error.

This is fully analyzed by Symantec \cite{tls-proxy} presenting a theoretical implementation model for TLS 1.3 traffic interception.

\subsubsection{Abusing the CA infrastructure}
The Certificate Authority infrastructure is built on trust, strong in theory, but in many cases CAs have been caught cheating in the wild \cite{ca-abuse}:
\myblockquote{We are now witnessing occasions in which digital certificates are being spoofed or hijacked and then sold on the black market.
  Attackers have been successful in misleading trusted CAs, including Verisign, Comodo, and DigiNotar, into issuing fraudulent certificates for websites, such as Microsoft and Google.
  In September 2011, stories surfaced involving DigiNotar, a Dutch “default‐trusted” root CA. Apparently, the CA may have been compromised maybe as early as 2009; 531 false certificates had been issued including certificates for domains such as Google, Facebook, Microsoft, Twitter, Mozilla, Mossad, MI6, and the CIA.}
These attacks raised concerns and new measures appeared, Google has been pushing for CT, \textit{Certificate Transparency}, introducing transparency logs, in which all the issued certificates appear, to ensure that no rogue certificate is emitted.
This is mostly a retrospective safety measure, losing the CA authority status is a good obstacle to anyone trying to abuse the system, but does not guarantee safety a priori.

\subsubsection{Hash collision}
A practical exploit for MD5 \cite{MD5-collision} has already been executed in 2005, raising many concerns.
By now both MD5 and SHA-1 algorithms have been deprecated, many browsers and applications will display warnings or refuse connection when presented with these hashes.

The practical solution found by the researchers was to alter specific bits of the hard to factor moduli to present identically signed certificates and forge a certificate that shared the same signature, hence being verified by a CA, but holding different private keys, allowing an eavesdropper to intercept traffic.

\begin{figure*}[h!]
  \centering
  \includegraphics{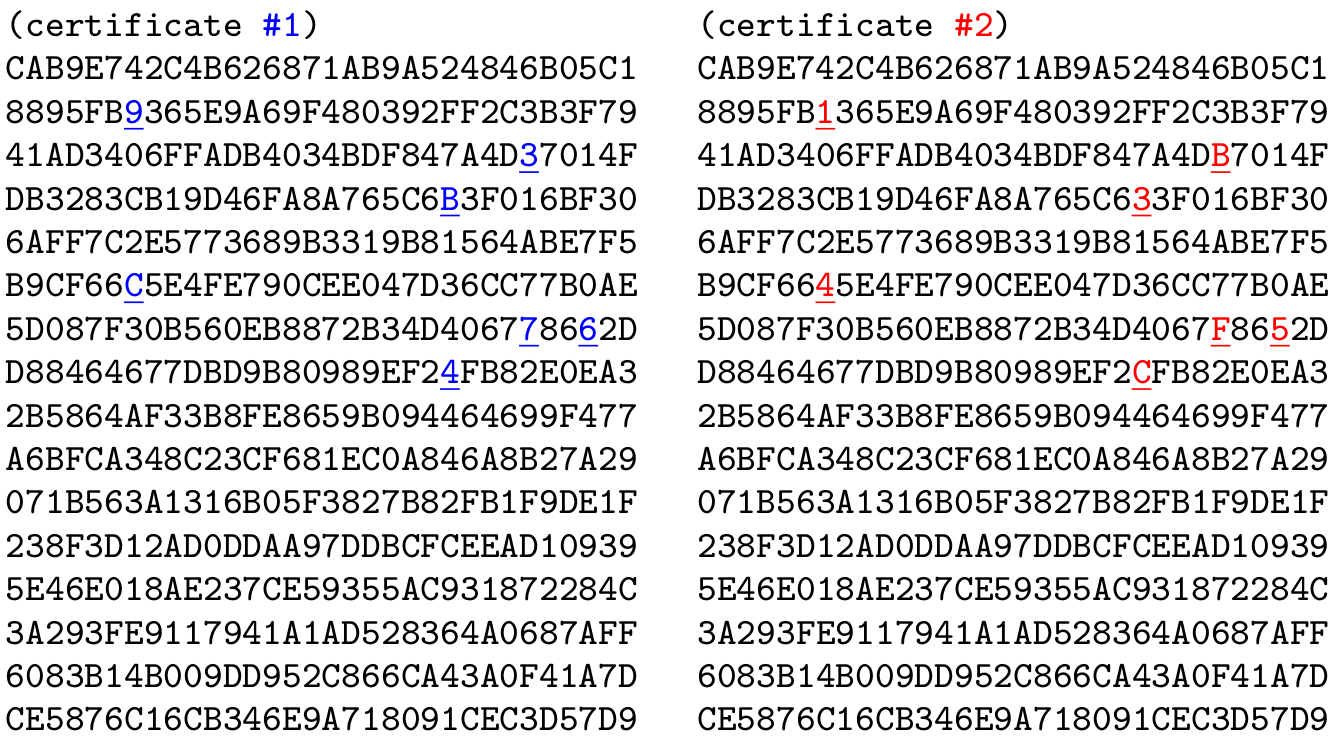}
  \caption{MD5 colliding certificates' public key moduli \cite{MD5-collision}}
  \label{img-colliding-certs}
\end{figure*}

\subsubsection{Weak cipher suites}
Many TLS cipher suites have been compromised \cite{weak-cipher}: BEAST (on CBC encryption in TLS 1.0), Lucky13 (on CBC padding in TLS 1.2), POODLE (on CBC padding in SSL 3.0), RC4 NOMORE (on keystream biases in RC4), FREAK (on export-grade RSAkeys), Logjam (on export-grade Diffie-Hellman groups), and DROWN (on RSA-PKCS\#1v1.5 encryption in SSL 2.0).

Leaving these chiper enabled in a webserver or client is bad practice, but widely done not to drop connection for older clients.
Many of this scenarios are vulnerable to \textit{downgrade attacks} by exploiting protocol flaws (e.g. Logjam, SLOTH), and flawed cipher suites should be disabled entirely.

In 2016 researchers were able to exploit 64-bit block cipher \cite{weak-cipher} intercepting few hundred gigabytes of data and using a birthday paradox attack on it.

\subsubsection{Compression attacks}
Starting with CRIME, \textit{Compression Ratio Info-leak Made Easy}, in 2012 a series of attack against TLS compression have been done. An attacker that will be able to induce a client to perform some specifically crafted requests will be able to use the size of the response as indicator of the content of the request.
This is especially true with \textit{CSRF tokens} as demonstrated by \textit{BREACH} attack \cite{BREACH} that will reveal secrets repeated in the body in less than a minute.

This types of attacks do not rely on crypto itself, as they can be used against stream ciphers as well as block ciphers, rather on implementation itself, thus being much more insidious.

\subsection{Structure}
The tunnel is designed to protect the data since the first API call because username, registration, and password recovery are critically sensitive information that if the protection started with SRP would not be encrypted.
Therefore, an ECDHE exchange is performed on landing to mediate an encryption key.
In detail this exchange goes as follows:
\begin{enumerate}
  \item the client requests the server key $Q$ with a \texttt{GET} request at \texttt{/tunnel/key};
  \item the server replies with a signed $Q$ point on the \texttt{curve25519} and the server's signing PGP key;
  \item the client verifies the PGP key against a Merkle tree for key transparency;
  \item the client generates the public parameter $V$ and the secret $Z$, submitting $V$ at \texttt{POST /tunnel/key};
  \item the server calculates the secret $Z$, and returns a newly generated session \texttt{UID};
  \item client and server hash $Z$ with the PGP key fingerprint to obtain a session key.
\end{enumerate}
From now on, all the traffic is going to have only the headers necessary for transmission and the \texttt{UID} in clear to allow the decryption.
The private signing PGP key is meant to be kept airgapped, so that a rogue server will not allow an attacker to know the private signing key, that can be used to re-sign a new $Q$.
This procedure has the drawback that it requires an administrator to rotate them manually.

When the log-in happens SRP generates a secret $K$ that can be used for the tunnel:
\begin{enumerate}
  \item the client sends the server the username with a \texttt{POST} request at \texttt{/auth/info};
  \item the server replies with an ephemeral challenge $S$, a modulus $M$, and a salt;
  \item the client computes the ephemeral parameter $C$, the secret $K$, and $P_C = \H(C,S,K)$ as authentication proof, then sends $C$ and $P_C$ at \texttt{POST /auth};
  \item the server then calculates $K$ using $C$, verifies $P_C$, generates $P_S$ to prove its identity to the client, and sends $P_S$ back;
  \item the secret $K$ is finally hashed by both parties and used as session key;
\end{enumerate}
As it can be observed the secret $K$ is generated with a personalised challenge for every user, using their own password, hence making it very difficult to crack this key on an extended scale.
For every user and connection the discrete logarithm problem must be solved, without giving a single failure point as the factorization of a public key.

It is worth underlining that an attacker could save the traffic for a long time to crack until it is able to find the password to crack it all at once, but the secret $K$ cannot be derived just knowing the password, hence preventing this scenario.
In that case a new session can be created, but no information can be derived onto a past or existing one.

The cipher suite Curve25519 with AES-128 was selected because of its robustness and convenience: the algorithms are already present into the OpenPGP standard \cite{RFC4880}, making implementation on clients easier, no additional library is necessary.

To make the tunnel more robust and to avoid birthday attacks the keys are refreshed after a set amount of time, enforced by the server with an error. The client is then allowed to use the old key just to refresh the session.
Since this exchange is encrypted the only information leaked is the timing and size of the packets, making it very difficult for an attacker to guess when is this happening.

\begin{figure*}[ht]
  \centering
  \begin{tikzpicture}
  \node at (-2, 3) {User actions};
  \draw[line width=0.40mm] (0, 0) rectangle (11, 2) node[pos=0.5] {\Large Tunnel};
  \draw[->, line width=0.50mm, dashed] (0, 4) -- (0, 2.05) node[midway, sloped, yshift=0.3cm, xshift=-0.3cm] {Landing};
  \draw[->, line width=0.50mm, dashed] (2, 4) -- (2, 2.05) node[midway, sloped, yshift=0.3cm, xshift=-0.5cm] {Login};

  \node at (-2, -1) {Exchanges};
  \draw[->, line width=0.50mm, dashed] (0, -2) -- (0, -0.05) node[midway, sloped, yshift=-0.3cm, xshift=-0.4cm] {ECDHE};
  \draw[->, line width=0.50mm, dashed] (2, -2) -- (2, -0.05) node[midway, sloped, yshift=-0.3cm, xshift=-0.6cm] {SRP};
  \draw[->, line width=0.50mm, dashed] (6, -2) -- (6, -0.05) node[midway, sloped, yshift=-0.3cm, xshift=-0.4cm] {ECDHE};
  \draw[->, line width=0.50mm, dashed] (10, -2) -- (10, -0.05) node[midway, sloped, yshift=-0.3cm, xshift=-0.4cm] {ECDHE};

  \draw[<->, line width=0.50mm, blue] (2.1, -0.9) -- (5.9, -0.9) node[midway, above] {Key refresh timeout};

  \draw[-, line width=0.30mm, dashed] (7, 2.5) -- (8, 2.5);
  \draw[->, line width=0.30mm] (8, 2.5) -- (11, 2.5) node[midway, above, xshift=-0.5cm] {Time};

  \node at (13, 0) {~};
\end{tikzpicture}
  \caption{Schematic tunnel structure}
  \label{fig-tunnel-schema}
\end{figure*}
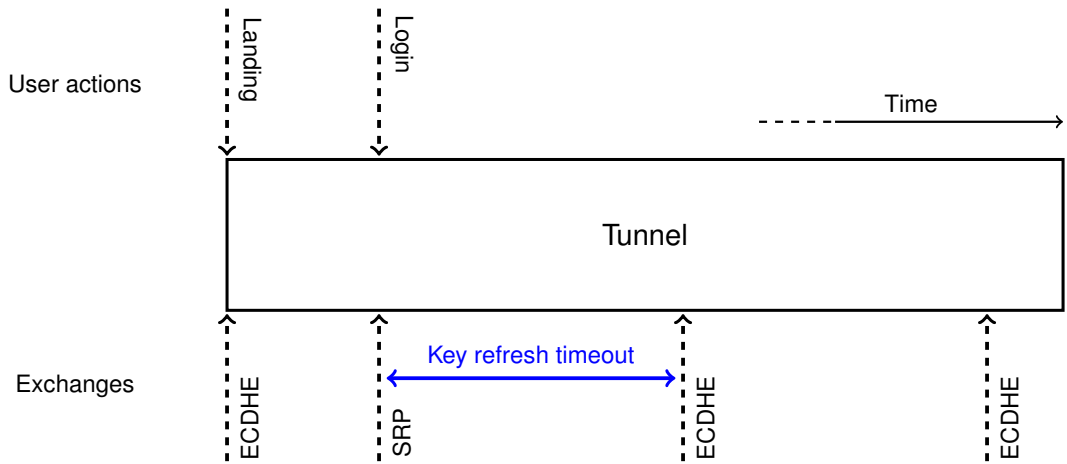

Figure \ref{fig-tunnel-schema} visualises this mechanism, in particular the user actions and the corresponding key exhanges.
It is important to understand that since exchanges are performed within a tunnel, to fully compromise the transferred information both exchanges must be broken, as even cracking the shared ECDHE $Q$ public parameter will not provide information after the SRP exchange, keeping the users' keys safe.

Once exchanges are performed, \texttt{AES-128-GCM} is used as symmetric encryption, as it is authenticated and fast, taking advantage of hardware acceleration on the server and webcrypto on the compatible clients.

\begin{figure*}[ht]
  \centering
  \begin{tikzpicture}
  \def\hrect {(0,0) rectangle (3,-1.5)}
  \def\brect {(0,-1.5) rectangle (3,-4)}
  \begin{scope}
    \draw \hrect node[pos=0.5] {Headers};
    \draw \brect node[pos=0.5] {Payload};
  \end{scope}

  \draw[->, line width=0.80mm] (3.2, -2) node[above,xshift=1cm] {Encryption} -- (5.3,-2);

  \draw[line width=0.40mm] (5.5, 0.5) rectangle (9.5, -4.5) node[pos=0.5, yshift=2.8cm] {\large Tunnel};
  \begin{scope}[shift={(6cm,0cm)}]
    \draw \hrect node[text width=2cm, align=center, pos=0.5] {Unencrypted\\headers};
    \draw \brect;
    \draw (0.2, -1.7) rectangle (2.8, -2.7) node[text width=2cm, align=center, pos=0.5] {Encrypted\\headers};
    \draw (0.2, -2.9) rectangle (2.8, -3.8) node[pos=0.5] {Payload};
  \end{scope}

  \draw[->, line width=0.80mm] (9.7, -2) node[above,xshift=1cm] {Decryption} -- (11.8,-2);

  \begin{scope}[shift={(12cm,0cm)}]
    \draw \hrect node[pos=0.5] {Headers};
    \draw \brect node[pos=0.5] {Payload};
  \end{scope}
\end{tikzpicture}
  \caption{Schematic tunnel packet transmission}
  \label{fig-packet-structure}
\end{figure*}
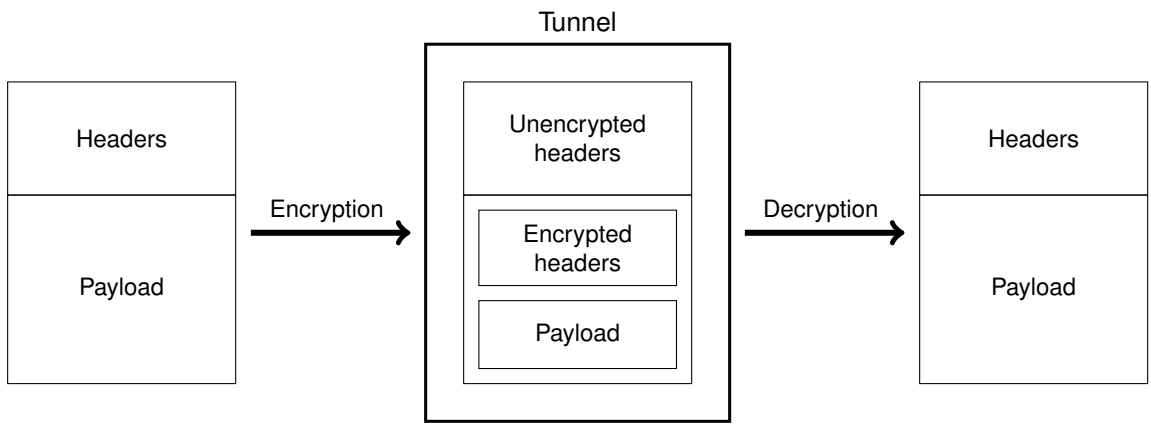

In figure \ref{fig-packet-structure} is represented the packet encryption and decryption mechanism: a plain \texttt{RFC 791} packet is encrypted before leaving the client. The necessary headers are kept unencrypted, and the rest is encrypted with the same structure provided by the standard: \texttt{CRLF} to separate each header and two times \texttt{CRLF} to separate the headers from the body. In this way all types of packet can be easily transmitted, whether they are plaintext, \texttt{JSON}, or binary data.

To prevent \textit{replay attacks} the packet is equipped with a nonce and a timestamp, a packet is allowed 120s to get to the server, after which is no longer considered valid.
In this timeframe the packets must always have a unique nonce, so that an attacker sending twice the same packet will have the second bounced before it is processed.

In table \ref{table-packet-structure} the precise structure of a packet is outlined. The encrypted part consists of the first 16 bytes of \textit{initialisation vector} (IV), followed by the \texttt{AES-128-GCM} encrypted data and its 12 byte authentication tag.
In this way we ensure that data cannot be modified by altering the IV or by tampering with the ciphertext.

The system is built to handle cookies both in the encrypted and unencrypted subsection, as the webclient uses \textit{secure cookies} that can not be accessed from the javascript application, hence making it impossible to encrypt them.
Since cookies store only the access and refresh tokens, that are associated with an encrypted session, they are useless without the encryption key, so can be transmitted in clear without any security trade-off.

\begin{table}[h!]
  \centering
  \def\arraystretch{1.8}
  \begin{tabular}{|c|c|c|}
\hline
\begin{tabular}[c]{@{}c@{}}Unencrypted\\ (Headers)\end{tabular}             & \multicolumn{2}{c|}{\begin{tabular}[c]{@{}c@{}}Necessary headers for \\ transmission and decryption\end{tabular}} \\ \hline
\multirow{4}{*}{\begin{tabular}[c]{@{}c@{}}Encrypted\\ (Body)\end{tabular}} & \multicolumn{2}{c|}{Initialisation Vector (IV)}                                                                   \\ \cline{2-3}
                                                                            & Headers              & \begin{tabular}[c]{@{}c@{}}Timestamp, nonce, \\ session, cookies*\end{tabular}             \\ \cline{2-3}
                                                                            & Payload              & \begin{tabular}[c]{@{}c@{}}Body of request \\ or response\end{tabular}                     \\ \cline{2-3}
                                                                            & \multicolumn{2}{c|}{Authentication tag}                                                                           \\ \hline
\end{tabular}
  \caption{Tunnel packet structure}
  \label{table-packet-structure}
\end{table}

Client and server, when decrypting packets will join the headers to re-create the orginal packet, in case of conflicting headers the encrypted ones will prevail.
This applies to cookies too, since they are in practice transmitted as a header.
A limitation of this method is the unability to send both, encrypted and unencrypted cookies, for instance the aforementioned case of the secure cookies.

\section{Implementation}
\subsection{Backend implementation}
\label{be-implementation}
This project is composed of several parts, an API controller, a \texttt{Session} model, and a \textit{middleware}.
The controller provides a three RESTful routes:
\begin{itemize}
  \item \texttt{GET /tunnel/key} to request the signed public parameter $Q$, a point on the curve25519, to perform the ECDHE;
  \item \texttt{POST /tunnel/key} to submit the ECDHE exchange and issue a new tunnel session, identified by an UID;
  \item \texttt{PUT /tunnel/key} to submit an ECDHE exchange to refresh the tunnel when it is expired.
\end{itemize}

This API relies on the \texttt{Session} model to store the derived encryption key and expiration, identified by the unique identifier \texttt{UID}.
When a client connects through the tunnel an encrypted session is created, that cannot in any case be converted to an unencrypted one.
A session is uniquely tied with an oauth AccessToken and RefreshToken through the UID, this means that if the encryption key is lost then the user must login again and a new session must be established.

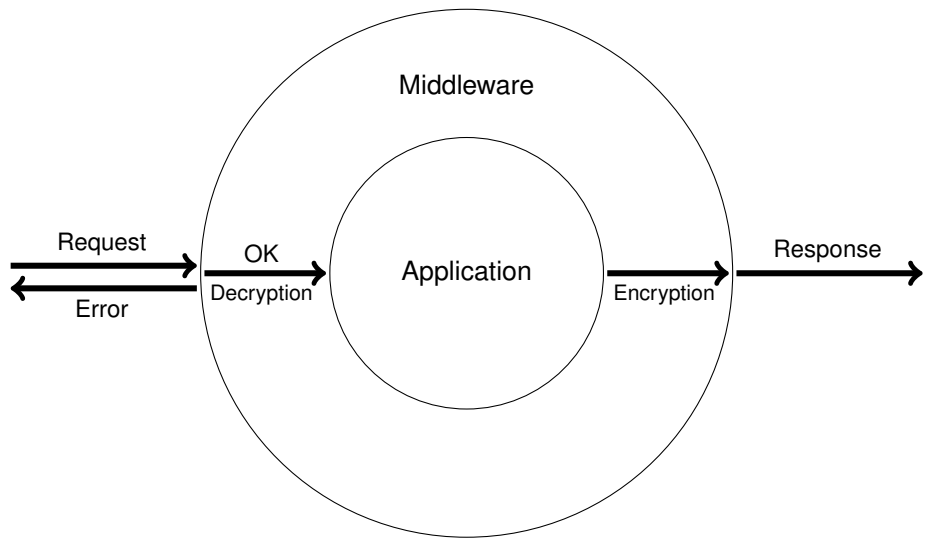
\begin{figure*}[h!]
  \centering
  \begin{tikzpicture}
  \draw (0,0) circle (1.8cm) node[pos=0.5] {\large Application};
  \draw (0,0) circle (3.5cm) node[pos=0.5, yshift=2.5cm] {\large Middleware};
  \draw[->, line width=0.80mm] (-6, 0.1) node[above,xshift=1.2cm] {Request} -- (-3.55, 0.1);
  \draw[<-, line width=0.80mm] (-6, -0.2) node[below,xshift=1.2cm] {Error} -- (-3.55, -0.2);
  \draw[->, line width=0.80mm] (-3.45, 0) node[above,xshift=0.75cm] {OK} node[below,xshift=0.75cm] {\small Decryption} -- (-1.85, 0);
  \draw[->, line width=0.80mm] (1.85, 0) node[below,xshift=0.75cm] {\small Encryption} -- (3.45, 0);
  \draw[->, line width=0.80mm] (3.55, 0) node[above,xshift=1.2cm] {Response} -- (6, 0);
\end{tikzpicture}
  \caption{Schematic middleware representation}
  \label{fig-middleware}
\end{figure*}

Once an encrypt session is established all requests to the API will be routed through \texttt{POST /tunnel/data}, to be then handled by a middleware, schematically represented in figure \ref{fig-middleware}.
It is executed as an outer shell of the application: it intercepts the requests, decrypts and verifies the validity of the packet, and if any of the parameters is wrong it will return an error to the client. If the server is able to decrypt the request the response or error is encrypted, otherwise it is unencrypted.

After verifying the previosly discussed security features, such as nonce and timestamp, the middleware creates a new request with the unencrypted route, parameters and payload, ready to be submitted to the router.
This component is run as an early middleware, allowing it to intercept the request almost before any parsing is done.
The API, being written in PHP, relies on some pre-processing that unfortunately is not available within the encrypted tunnel and is to be done manually, specifically the superglobals: \texttt{\$\_GET}, \texttt{\$\_POST}, \texttt{\$\_COOKIES}, \texttt{\$\_SERVER}, and \texttt{\$\_FILES}.

Encrypted headers, once decoded, are added to the new request merging them with the \texttt{\$\_SERVER} superglobal.
Since cookies are also a header, they are specifically parsed by a class according to RFC 6265 \cite{RFC6265}, subsection 5.4.

The API mostly handles \texttt{JSON} encoded requests, with some exceptions, such as attachment submission or file upload.
The discriminant factor is the \texttt{Content-Type} header, when it is either \texttt{multipart/form-data} or \texttt{application/x-www-form-urlencoded} the request is additionally parsed by a class built upon RFC 2388 \cite{RFC2388}. Many edge cases and the necessity to parse data coming form a moltitude of different clients and browsers made this task particularly intricated.

\subsection{Frontend implementation}
\subsubsection{Client threats}
The tunnel has been designed to address the \textit{threat model}, given that the front end code is not flawed.
On application clients, like the Android or iOS app, as well as the bridge and the VPN apps this is easily addressable, while for the webclient extra safety measures must be implemented, to verify the JS app integrity.
This project is being developed parallel to a code integrity verification system, that will provide trust on the application layer.

\paragraph{Key Verification}
To verify the signature on the public parameter $Q$ of the ECDHE exchange a public signing key is available in ProtonMail's key verification system.
This technology is based upon COINKS \cite{COINKS}, an end-user key verification service capable of integration in end-to-end encrypted communication systems.
The clients will have to first retrieve the key from the public key API, then verify it against the Merkle tree as described in the paper, to then verify the signature.

\subsubsection{Migration}
Clients must be allowed to communicate in both the encrypted and unencrypted form; since it will require time to implement this feature the tunnel will not be enforced on all connections.
If at the beginning of a session an ECDHE key is negiotiated between the two parties then the connection will be stored as encrypted and it will not be possible to communicate without encryption using that specific identifier, as it could be used as a downgrade attack.

Sessions are in fact divided in encrypted and unencrypted and it won't be allowed to switch between the two states.
Disabling encryption on a session can be overridden by the server, for instance when testing, since the developer tools in any browser will just show unintelligible blobs of data being sent to \texttt{POST /tunnel/data}, and using any API development tool will turn out to be overly complicated.

\subsection{Testing}
To test the API implementation several approaches have been followed: unit tests for the single modules, functional tests for the API and middleware, end-to-end tests for a bigger-picture testing, and a proxy for manual testing.

Unit and functional tests have been developed to work on \textit{Continuous Integration} (CI), and be automatically performed, while end-to-end must be run with a fully functioning HTTP server to test the middleware's interaction with the requests.

All of these tests are designed to test several API functionalities, such as key exchanges and renegotiation, and the middleware functionalities, from the basic encryption and decryption to the edge cases of encrypted cookies and multipart requests.

\subsubsection{Test proxy}
To perform manual testing of the tunnel a NodeJS proxy has been developed, nicknamed \texttt{NodeProxy}, that provides a compatibility interface between the existing clients and the API with the tunnel enabled.

To mediate this interface the test proxy:
\begin{enumerate}
  \item performs the key exchange using the OpenPGPJS library, directly accessing the newly-inserted \texttt{genPublicEphemeralKey} function;
  \item initiates a session and displays the current session UID and encryption key, taking care of encryption and decryption of all traffic;
  \item on login intercepts the SRP params and performs a MITM attack and updates the encryption key;
  \item when key refresh is necessary it locks and queues all the requests in an await status, to perform a new handshake.
\end{enumerate}

This proxy is completely transparent to the clients, it is needed just to alter the API address to match the proxy's.

In order to perform SRP transparently this tool is requires the user's password and features a full implementation of an inline MITM attack. The proxy:
\begin{enumerate}
  \item detects a log-in attempt by listening to the API calls, when a call to \texttt{POST /auth/info} a verifier and a challenge are generated, emulating an SRP server;
  \item forwards the signed modulus in the server's response, and alters the ephemeral challenge;
  \item detects a call to \texttt{POST /auth} and generates a client ephemeral and proof;
  \item responds to the client with a newly-generated server proof and confirms the login.
\end{enumerate}

\begin{figure*}[ht]
  \centering
  \begin{tikzpicture}
  \node at (-6, 0) {\large Client};
  \node at (0, 0) {\large Proxy};
  \node at (6, 0) {\large Server};

  \draw[line width=0.5mm] (-2.05,0) -- (-2.05,-21.1);
  \draw[line width=0.5mm] (-3.95,0) -- (-3.95,-21.1);

  \draw[line width=0.5mm] (2.05,0) -- (2.05,-21.1);
  \draw[line width=0.5mm] (3.95,0) -- (3.95,-21.1);

  \draw[->, line width=0.3mm] (-3.8,-1) -- (-2.2,-1) node[midway, above] {\small username};
  \draw[->, line width=0.3mm] (2.2,-1) -- (3.8,-1) node[midway, above] {\small username};

  \node at (6, -1) {Generates $s_s$};
  \node at (6, -1.7) {Fetches $v_s$, $salt_s$};
  \node at (6, -2.4) {$k = \H(M, g)$};
  \node at (6, -3.1) {$S_s = g^{s_s} + kv_s$};

  \draw[<-, line width=0.3mm] (2.2,-3.1) -- (3.8,-3.1) node[midway, above] {\small $S_s$, $salt_s$, $M$};

  \node at (0, -3.1) {Generates $c_p$};
  \node at (0, -3.8) {$k = \H(M, g)$};
  \node at (0, -4.5) {$x_s = \H(p, salt_s)$};
  \node at (0, -5.2) {$C_p = g^{c_p}$};
  \node at (0, -5.9) {$u_s = \H(S_s, C_p)$};
  \node at (0, -6.6) {$g^{s_s} = S_s - kg^{x_s}$};
  \node at (0, -7.3) {$K_{ps} = (g^{s_s})^{c_p+u_sx_s}$};
  \node at (0, -8.0) {$P_{Cp} = \H(S_s, C_p, K_{ps})$};

  \node at (0, -8.7) {Generates $s_p$, $salt_p$};
  \node at (0, -9.4) {$v_p = g^{\H(p, salt_p)}$};
  \node at (0, -10.1) {$S_p = g^{s_p} + kv_p$};

  \draw[<-, line width=0.3mm] (-3.8,-10.1) -- (-2.2,-10.1) node[midway, above] {\small $S_p$, $salt_p$, $M$};

  \node at (-6, -10.1) {Generates $c_c$};
  \node at (-6, -10.8) {$k = \H(M, g)$};
  \node at (-6, -11.5) {$x_c = \H(p, salt_p)$};
  \node at (-6, -12.2) {$C_c = g^{c_c}$};
  \node at (-6, -12.9) {$u_c = \H(S_p, C_c)$};
  \node at (-6, -13.6) {$g^{s_p} = S_p - kg^{x_c}$};
  \node at (-6, -14.3) {$K_{cp} = (g^{s_p})^{c_c+u_cx_c}$};
  \node at (-6, -15.0) {$P_{Cc} = \H(S_p, C_c, K_{cp})$};

  \draw[->, line width=0.3mm] (-3.8,-15.0) -- (-2.2,-15.0) node[midway, above] {$C_c$, $P_{Cc}$};

  \node at (0, -15.0) {$u_c = \H(S_p,C_c)$};
  \node at (0, -15.7) {$K_{pc} = (C_c v_p^{u_c})^{s_p}$};
  \node at (0, -16.4) {$P_{Cc} \stackrel{?}{=} \H(S_p, C_c, K_{pc})$};
  \node at (0, -17.1) {$P_{Sp} = \H(S_p, P_{Cc}, K_{pc})$};

  \node at (0, -17.8) {Fetches $C_p$, $P_{Cp}$};

  \draw[->, line width=0.3mm] (2.2,-17.8) -- (3.8,-17.8) node[midway, above] {$C_p$, $P_C$};

  \node at (6, -17.8) {$u_s = \H(S_s,C_p)$};
  \node at (6, -18.5) {$K_{sp} = (C_p v_s^u)^{s_s}$};
  \node at (6, -19.2) {$P_{Cp} \stackrel{?}{=} \H(S_s, C_p, K_{sp})$};
  \node at (6, -19.9) {$P_{Ss} = \H(S_s, P_{Cp}, K_{sp})$};

  \draw[<-, line width=0.3mm] (2.2, -19.9) -- (3.8, -19.9) node[midway, above] {$P_{Ss}$};

  \node at (0, -19.9) {$P_{Ss} \stackrel{?}{=} \H(S_s, P_{Cp}, K_{ps})$};
  \node at (0, -20.6) {Fetches $P_{Sp}$};

  \draw[<-, line width=0.3mm] (-3.8, -20.6) -- (-2.2, -20.6) node[midway, above] {$P_{Sp}$};

  \node at (-6, -20.6) {$P_{Sp} \stackrel{?}{=} \H(S_p, P_{Cc}, K_{cp})$};

  \draw [line width=0.3mm,decorate,decoration={brace,amplitude=10pt}] (-1.5,-21.4) -- (-4.5,-21.4);
  \node at (-3, -22.1) {$K_{cp} = K_{pc} = g^{(c_c+u_cx_c)s_p}$};

  \draw [line width=0.3mm,decorate,decoration={brace,amplitude=10pt}] (4.5,-21.4) -- (1.5,-21.4);
  \node at (3, -22.1) {$K_{sp} = K_{ps} = g^{(c_p+u_sx_s)s_s}$};
\end{tikzpicture}
  \caption{SRP MITM attack, inline, as implemented in \texttt{NodeProxy}.
  {\color{darkgray} $M$ is the fixed Modulus, $g$ the generator, $p$ is the user's password, $v_s$ is the server's password verifier.}}
  \label{fig-srp-mitm}
\end{figure*}
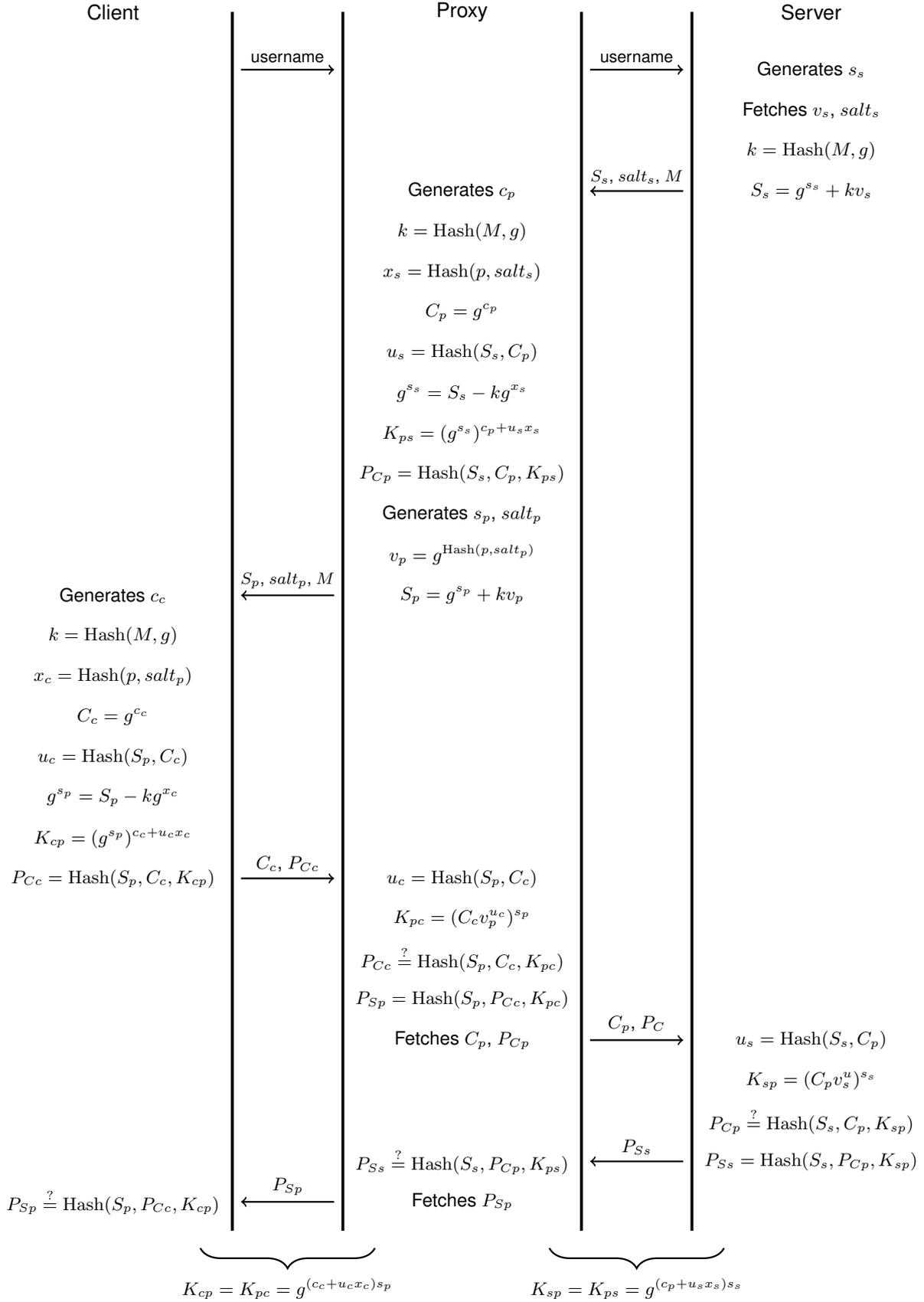

This attack, illustrated in figure \ref{fig-srp-mitm}, is possible only by knowing the user's password, and required writing a JS implementation of an SRP server, previously not available.

\section{Conclusions}
The aim of this project is to guarantee the security and authenticity of all the information shared between the client and the server by encrypting every request to the server, before transmitting it over TLS.
This is achieved by studying and addressing TLS's main weaknesses, thus significantly raising the complexity of an attack.

In most cases TLS is being systematically intercepted by proxies, that would be prevented to access and alter the transmitted information, thus only being capable to block the user from connecting ProtonMail's services without being able to tamper with the data.
Malicious TLS compromise is very alarming because it is a critical issue with a very little footprint, as it is usually a relatively elaborate attack,
that requires a targeted intent.
In this case cracking the second layer of encryption renders the attack much more difficult, either by altering the client code or by cracking another set of encryption keys.

The tunnel secures the packets by encrypting them with a symmetric cipher, keeping public only the session identifiers necessary to fetch the correct decryption key.
In order to mediate a session key, two different key exchanges are used: the first using elliptic curves immediately on landing, the second taking advantage of the residual secret from the password exchange protocol.
Login is therefore a critical security step for the tunnel: assuming an attacker does neither know the password verifier, nor the password itself, the SRP exchange is immune to MITM attacks.
In fact, the tunnel not only protects the transmitted data, but ensures full authenticity of the connection, guaranteeing that every non-compromised client implementation is communicating exclusively with ProtonMail's authorized servers.

Nevertheless, this does not mean that TLS is broken, and this implementation does not supply a replacement to it.
The tunnel can indeed be used to transfer data over an insecure channel, but it is designed to co-operate with TLS and remedy to the most critical aspects of it.
Moreover, this project is not meant to be an implementation standard as it does not fit most use modes of TLS.
In fact, the project derives from a further need of data protection for high-confidentiality applications and is tailored onto ProtonMail's specific case.

\section*{Acknowledgments}
I would like to thank Francisco Vial for the help reviewing this paper.

\end{document}